\documentclass[prl,aps,twocolumn,,showpacs,floatfix]{revtex4}

\usepackage{graphicx}
\usepackage{dcolumn}
\usepackage{bm}

\begin{document}

\title{Luttinger Liquid at the Edge of a Graphene Vacuum}

\author{H.A. Fertig$^1$ and L. Brey$^2$}

\affiliation{\centerline {1. Department of Physics, Indiana University, Bloomington, IN 47405}
\\ 2. Instituto de Ciencia de Materiales de
Madrid (CSIC),~Cantoblanco,~28049~Madrid,~Spain}
\

\begin{abstract}
We demonstrate that an undoped two-dimensional carbon plane (graphene) whose
bulk is in the integer quantum Hall regime supports a non-chiral
Luttinger liquid at an armchair edge.  This behavior arises due to
the unusual dispersion of the non-interacting edges states, causing
a crossing of bands with different valley and spin indices at the edge.
We demonstrate that this stabilizes a domain wall structure with
a spontaneously ordered phase degree of freedom.  This coherent
domain wall supports gapless charged excitations, and has a power law
tunneling $I-V$ with a non-integral exponent.  In proximity to a
bulk lead, the edge may undergo a quantum phase transition between the Luttinger
liquid phase and a metallic state when the edge confinement is
sufficiently strong relative to the interaction energy scale.

\end{abstract}


\maketitle

{\it Introduction --} Two-dimensional carbon sheets
\cite{novoselov}, known as graphene, are emerging as one of the
most exciting new systems supporting the quantum Hall effect. This
material is different than more standard two-dimensional electron
gases (2DEGs) because in the absence of a magnetic field, the
single particle spectrum is linear in the vicinity of two
inequivalent points in the Brillouin zone. The low-energy states
near these points are described by the Dirac equation \cite{ando},
and in a strong magnetic field, the quantum Hall steps that emerge
are shifted relative to standard 2DEGs \cite{zheng,zhang}.  This
effect arises because the spectrum of the Dirac equation with a
magnetic field has doubly degenerate Landau levels (LL) for each
spin, with one pair at zero energy, half of which are filled in
the nominally undoped situation. Thus filling the Landau states to
reach the first quantum Hall plateau requires only half as many
electrons as needed to reach the subsequent plateaus, shifting the
step pattern.

In an undoped standard 2DEG system, there is little interesting
electron physics because the filled valence states are far below
the chemical potential in practical situations. By contrast, the
partially filled LL at zero energy in graphene allow for
interesting low-temperature physics even in this nominal
``vacuum''.  When interactions are included, the half-filled zero
energy states represent a multicomponent system, which in the
absence of spin or valley splitting potentials spontaneously
polarizes due to exchange \cite{multi_review}.  In practice, it is
spin-polarized in the bulk due to the Zeeman splitting.  As we
show below, the vacuum is thus a quantum Hall ferromagnet, with an
associated low-energy spin-wave.

\begin{figure}
  \includegraphics[clip,width=9cm]{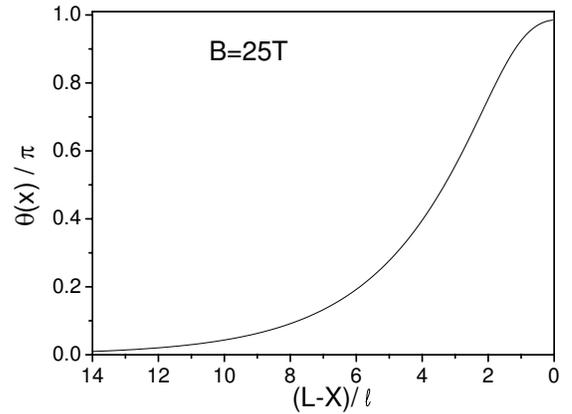}
  \caption{
Example of a domain wall configuration.  See text.}
 \label{Fig1}
\end{figure}

In addition to these surprising bulk properties, graphene also has
an unusual edge structure even in a non-interacting picture: the
lowest LL  has half as many edge states as the higher LLs
\cite{brey1}. In this work, we demonstrate a remarkable effect
when the edge structure and the quantum Hall ferromagnetism are
both taken into account. Under appropriate circumstances undoped
graphene forms a a coherent domain wall (DW) between the spin polarized state in
the bulk and an unpolarized region at the edge. The low-energy
theory of this DW has a U(1) symmetry with a Luttinger
liquid Hamiltonian \cite{mahan}. More specifically, the DW may be
described by a variational wavefunction of the form
\begin{equation}
|\Psi> = \! \! \prod _{X< L} \! \! \left (\!  \cos {\frac {\theta
(X)}{2} }\,
\!  + \! \sin {\frac {\theta (X)} {2}} \, e ^{i \phi} C ^{\dag}
_{- \downarrow   X} C_{+  \uparrow  X}\right  ) \! |{\rm Vac}\!>\,
\, \,, \label{wf}
\end{equation}
where $C ^{\dag} _{+  \uparrow  X}$ and $C ^{\dag} _{-  \downarrow
X}$ create electrons in the two relevant levels, $X$ is the
guiding center quantum number with allowed values up to the edge
at $L$, $|{\rm Vac} \! >$ denotes the bulk undoped graphene state
(i.e., vacuum) which is partially polarized since the two spin up
lowest LLs are fully occupied, and $\theta(X)$ and $\phi$ are
variational parameters.  An example of $\theta(X)$ found by
minimizing the effective energy functional is illustrated in Fig.
1.  The energy of the state is independent of $\phi$, indicating a
spontaneously broken symmetry in the DW  groundstate, with an
associated gapless collective mode which may be understood as
states in which $\phi$ has a spatial gradient
\cite{bilayer1,bilayer2,rene}.  Gradients in $\phi$ carry a charge
density, and a full $2\pi$ rotation contains a single electron
above the vacuum \cite{kane,falko}. Thus this system supports {\it
gapless} charged excitations.

This coherent DW  may be probed by tunneling into it from a bulk
metallic lead \cite{tunneling_expt}.  For a standard 2DEG, in the
undoped case the system trivially behaves as an insulator, and for
integer quantum Hall states one finds a metallic response
\cite{tunneling_expt}.  For the coherent DW, we expect a power law
tunneling $I-V$, with exponent determined by the pseudospin
stiffness and the strength of the confinement potential.  This is
quite different than standard quantum Hall edges, where the
tunneling exponent is thought to be set by the bulk filling factor
\cite{kaneandfisher}. By varying the strength of the
electron-electron interaction (for example, by a screening gate)
or the edge confinement potential, one can vary the tunneling
exponent, and in in principle may drive the system through a
quantum phase transition in which the tunneling perturbation
becomes relevant in the renormalization group sense.  This
presumably drives the system into a metallic state with a linear
$I-V$.

We now provide details of these results.

\begin{figure}
  \includegraphics[clip,width=9cm]{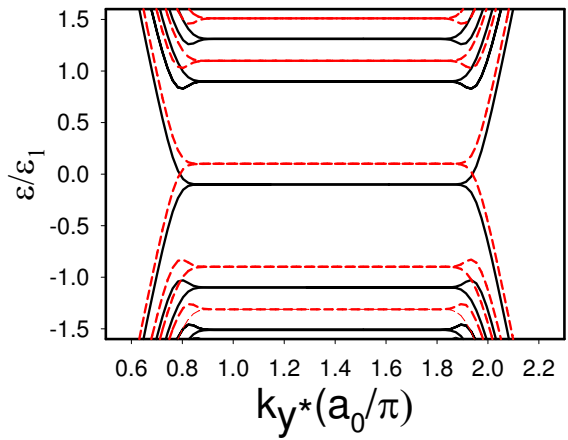}
  \caption{($Color$  $online$)
Energy bands for electrons in a graphene ribbon with armchair
edges in a magnetic field from tight binding calculations. $a_0=$ lattice constant,
unit of energy $\varepsilon_1=\sqrt{2}\gamma a_0/\ell$.
$k_y=X/\ell^2$ with $X$ the guiding center coordinate.
$B$=100T, ribbon width is 460\AA.}
 \label{Fig2}
\end{figure}

{\it Pseudospin Ferromagnetism in Graphene} -- The non-interacting
spectrum of a graphene ribbon in a magnetic field with armchair
edges is illustrated in Fig. 2 \cite{brey1}.  The eigenstates are
characterized by a  quantum number $X$ which specifies the center
of the wavefunction along the direction perpendicular to the edge,
and spin ($\sigma=\pm \frac{1}{2}$) and valley indices ($\tau=\pm
1$) for a total of four degenerate Landau bands inside the bulk of
the system.  In the undoped system, all the negative energy states
and two of the zero energy states \cite{com} are filled at zero
temperature. In what follows, we will ignore the LLs well away from
zero energy, since these are either completely filled or empty and
will not affect the physics described here.

Retaining just the four lowest Landau levels (LLLs) near zero energy,
apart from constant terms the Hamiltonian may be written as
\begin{eqnarray}
H & = &   \sum _{\tau, \sigma, X} \left [ -E_z \sigma C ^{\dag}
_{\tau \sigma X} C  _{\tau \sigma X}  + \Delta (X) \, \tau \,  C
^{\dag} _{\tau \sigma X} C  _{\tau \sigma X}  \right ] \nonumber \\
&+& \frac {N_{\phi} ^2} {2 S} \sum _{\sigma ,\sigma ',\tau ,\tau '
,\bf q} e ^{- \frac {q ^2 \ell ^2}{2}} \, V_{\bf q} \,
 \, \, \rho _{\tau,\sigma} (-{\bf q}) \, \, \rho _{\tau ',\sigma '} ({\bf q}) \, \, .
\end{eqnarray}

In this expression, $E_Z= g \mu _B B$ is the Zeeman coupling,
$\Delta(X)$ is the energy
splitting produced by the edge \cite{brey1},
$N_{\phi}/S$ is the number of flux quanta per unit
area passing through the system,
and
$V_{\bf q}= \frac {2 \pi e ^2} {\epsilon _0 q}$ is the Coulomb interaction.
Note that we have assumed an SU(4) symmetric form for the interaction.
Although not exact this should be an
excellent approximation for LLL states
provided the magnetic length is much larger than the lattice constant \cite{ando}.
Finally,
the LLL
density operators have the form
$
\rho _{\tau ,\sigma } ({\bf q}) = \frac {1}{N_{\phi}} \sum _{X} e ^{ - \frac {i}{2} q _x (2X+q_y)} C ^+ _{\tau
\sigma X} C  _{\tau \sigma X + q_y}.
$

The vacuum state (i.e., undoped groundstate) involves filling two
of the four LLL.  For $\Delta=E_z=0$, any two orthogonal states
will yield the same energy. In the bulk (where $\Delta=0$) the
Zeeman coupling breaks this symmetry, and in the Hartree-Fock
approximation the vacuum state may be written in the form
\begin{equation}
|{\rm Vac}> = {\prod }_{X} C ^{\dag} _{+  \uparrow  X} \, \,
C^{\dag}  _{- \uparrow  X  } \bigl|n<0\bigr>,
\end{equation}
where $\bigl|n<0\bigr>$ denotes the state with all the negative energy
levels filled.  Our vacuum is thus ferromagnetic.
The energy per particle of the ground state is
$
-E_z + \frac {1}{2}  \Sigma
$
with
$\Sigma =-\sqrt { \frac {\pi}{2}} \frac {e ^2} {\epsilon _0 \ell}$ the electron self-energy.
Note that this energy scale is much larger than
$E_z$, and that
the primary role of $E_z$ is to pick out which two states are actually
occupied.

The ferromagnetic vacuum supports four collective neutral excitations, which may be approximately generated by applying the
operators
\begin{equation}
\rho _{\tau , \tau ' \downarrow , \uparrow }  ({\bf q}) = \frac
{1}{N_{\phi}} {\sum} _{X} e ^{ - \frac {i}{2} q _x (2X+q_y)} C ^+
_{\tau \downarrow  X} C  _{\tau ' \uparrow  X + q_y}
\end{equation}
to $|{\rm Vac}>$. Each of these involves a spin flip, and two of them also include a
valley density wave. Because of the SU(4) symmetry of the interactions,
all four excitations are degenerate with
energy at long wavelengths
$\omega_0(q) \approx 2E_z+4\pi\rho_s q^2\ell^2$,
with $\rho_s=1/16\sqrt{2\pi}$ in units of $e^2/\epsilon_0\ell$, and
$\ell=\sqrt{c/eB}$ is the magnetic length.
This is precisely the form one expects for a ferromagnet in
the presence of a Zeeman coupling.

It is interesting to notice that due to the spin-charge coupling
inherent in a LL, doping the system could degrade the spin
polarization \cite{skyrme_lat}. Previous estimates \cite{paredes}
suggest that  $E_z$ is too large in the bulk for this to occur.
However, near the edge where the non-interacting levels cross, we
now demonstrate that such effects are crucial for obtaining a
correct description of the lowest energy states of the system.

{\it Coherent Domain Wall at the Edge} -- Fig. 2
illustrates the behavior of the non-interacting levels with 
armchair edges.
Because of the Zeeman splitting, a hole-like spin-up state
necessarily crosses an electron-like spin-down state
\cite{brey1,cross}. We note that the $\tau =+,-$ states near the
edge do not represent states purely in either valley, as they are
admixed by the boundary condition for an armchair edge
\cite{brey1}. Because of its effective spin stiffness, the
groundstate does not adhere to the lowest energy single particle
state, but rather generates a DW to make the spatial transition.
Eq. \ref{wf} describes such a DW when $\theta(X)$ passes from $0$
to $\pi$.  The energy of this state can be shown to have the form
\cite{cote2,unpub}
$$
E \simeq \pi \ell ^2 \rho _s  \sum _{X <L} \left ( \frac {d \theta} {d x} \right ) ^2+ \sum _{X<L}    \left [
\Delta (X) -E_z \right ] { \cos {\theta (X)}   } \label{functional},
$$
for an edge at $x=L$, provided $\theta(X)$ does not vary rapidly
on the scale of $\ell$. This sine-Gordon-like energy functional
may be straightforwardly minimized numerically \cite{unpub}. A
typical example of a DW solution is illustrated in Fig. 1.

Because the energy of $|\Psi>$ is independent of $\phi$, it is
immediately clear that this represents a broken symmetry state
which must support a gapless mode.  The DW is effectively a
one-dimensional easy-plane ferromagnet, which may be described by
an effective action
\begin{equation}
S_0 \! \! =  \! \! \! \int \! \! \! d \tau d y \! \left [ \frac
{\Gamma}{2}  m(y,\tau)^2 \! \! + \! \! \frac {\tilde{\rho}}{2}
\left ( \frac {\partial \phi }{\partial y } \right ) ^2 \! \! + \!
\! \! i m(y,\tau) \! \left ( \frac {\partial \phi }{\partial \tau
} \right )\! \right ] \, \, \, \label{action}
\end{equation}
where $\tau$ is imaginary time. The field $m$ has its origin in
the variable $\theta$, and may be understood as the position of
the center of the DW. The constants $\Gamma$ and $\tilde{\rho}$
may be estimated from $\theta(X)$ and $E_z-\Delta(X)$ using
spin-wave theory \cite{rene,unpub}. Details of this will be
published elsewhere.

Because $\theta(X)$ passes from 0 to $\pi$ over the width of the domain wall,
any configuration of $\phi(y)$ that passes from 0 to $2\pi$ represents
a state in which the sphere of allowed directions of an effective spinor
constructed from the two states $(+,\uparrow)$ and $(-,\downarrow)$ is
covered precisely once.  In this situation the state carries an excess or
deficit of one electron relative to the groundstate \cite{kane,falko}.
Moreover, the twist in $\phi$ may be spread out over the entire length of
the domain wall, so that charge may be introduced or removed with arbitarily
low energy, in contrast to the gap present for
charged excitations in the bulk.  As we now demonstrate, this has
important implications the tunneling into the edge of graphene vacuum.

\begin{figure}
  \includegraphics[clip,width=9cm]{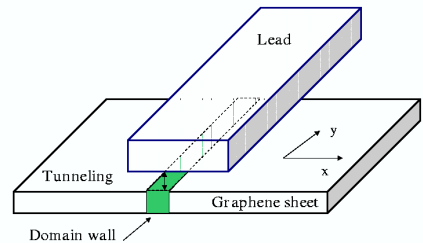}
  \caption{($Color$  $online$)
Geometry for tunneling into the coherent domain wall.}
 \label{Fig3}
\end{figure}

{\it Tunneling from a Metallic Electrode} -- The geometry for
tunneling into the DW is illustrated in Fig. 3.  We model the lead
as a simple Fermi liquid, with action
$$
S_{Lead}\! \! = \! \!  -\frac{1}{\beta}   \! \! \!
\sum_{\omega_n,n,k_z,k_x} \! \! \!
[E_n(k_z)-\mu+i\omega_n]a^*_{nk_yk_z}(\omega_n)a_{nk_yk_z}(\omega_n),
$$
where $\beta$ is the inverse temperature and $\mu$ the chemical potential.
The Grassman variable
$a_{nk_xk_z}$ corresponds to the
destruction operator for a lead electron in the $n$th
Landau level state $\Phi_{nk_yk_z}=\frac{1}{\sqrt{L_yL_z}}
e^{ik_yy+ik_zz}\varphi_n(x-k_y\ell^2)$ with $\varphi_n$ a
harmonic oscillator state, and
$E_n(k_z)=(n+1/2)\omega_c+k_z^2/2m$, $\omega_c=eB/mc$ being the
cyclotron frequency.  The tunnel coupling between the lead and
reservoir takes the form
$$S_{tun}\!=\!\frac{1}{\beta}\sum_{\omega_n}\sum_{nk_y} \! \int \! dy
\chi _{nk_y}(y)a_{nk_yk_z}^{*} (\omega_n) \psi(y,\omega_n) +
h.c.$$ where $\psi$ is the annihilation operator for an electron
in the domain wall, which we have taken to be located at $x=z=0$,
and
$\chi_{nk_y}(y)=\frac{t}{\sqrt{L_y}}e^{-ik_yy}\varphi_n(-k_y\ell^2)$.
Using standard bosonization \cite{mahan}, one may write the
fermion operator as $\psi(y,\tau) \sim
e^{-i\phi(y,\tau)}e^{i2\pi\int_{-\infty}^x dy^{\prime}
m(y^{\prime},\tau)}$.  After tracing out the lead degrees of
freedom, one finds the partition function function may be written
in the form $Z=\int{\cal D}^2a{\cal D}\phi{\cal D}m
e^{-S_0-S_{Lead}-S_{tun}} \propto \int{\cal D}\phi{\cal D}m
e^{-S_0-\tilde{S}}$, with
$$
\tilde{S}=-\frac{1}{\beta} \sum_{\omega_n}\sum_{nk_yk_z} \int
dy_1dy_2 \frac{\chi_{nk_y}(y_1)\chi ^*_{nk_y}(y_2)
\psi^*(y_1,\omega_n)\psi(y_2,\omega_n)}{i\omega_n-\mu+E_n(k_z)}.
$$
Without qualitatively changing the result, one can restrict the LL
sum to just the LLL.  By noting that $\sum_{k_y} \chi \chi ^*$ is
strongly peaked at $y_1\!=\!y_2$, $\tilde{S}$ may be recast in the
form
\begin{equation}
\tilde{S}=-t^2\int_0^{\beta}d\tau_1 d\tau_2\int dy
\psi^*(y,\tau_1)K(\tau_1-\tau_2)\psi(y,\tau_2).
\label{Sres}
\end{equation}
Taking the zero temperature limit, one may show \cite{unpub}
$K \sim 1/(\tau_1-\tau_2)$ for large $|\tau_1-\tau_2|$.

Our first question is whether the presence of $\tilde{S}$
qualitatively affects the state of the system; i.e., is it
a relevant operator.  A perturbative renormalization group (RG)
analysis may be applied to $\tilde{S}$ \cite{unpub}, leading
to the result
$$
\frac{dt^2}{d \ell} = -(\kappa-2)t^2
$$
with the anomalous dimension $\kappa=(x+1/x)/2$, and $x=4\pi
\sqrt{\tilde{\rho}/\Gamma}$. Estimates of $\tilde{\rho}$ and
$\Gamma$ using a spin-wave approach \cite{unpub} yield $\kappa
\approx 6.8,~6.0,$ and 5.3 for $B=$15T, 25T, and 45T,
respectively. This indicates that under usual conditions,
$\tilde{S}$ is irrelevant, and the DW remains in a Luttinger
liquid phase in spite of the coupling to the lead.  The physical
reason for this is that the pinning energy of the domain wall is
small compared to the stiffness of the phase angle, because the
Zeeman energy which ultimately creates the DW is quite small
compared to the electron-electron energy scale. This suggests that
enhancing the relative pinning energy can drive the system into a
state in which $\tilde{S}$ is relevant, perhaps by judicious use
of a gating geometry. In this situation the coupling to the
metallic lead becomes important in the low-energy physics, and
presumably a current injected into the domain wall will behave
metallically.

The irrelevance of $\tilde{S}$ under ambient conditions indicates
that we can treat it perturbatively.  The tunneling current is then
a convolution of the spectral functions for the lead and the
domain wall, separated in frequency by $eV$, with $V$ the potential
difference between the two systems \cite{mahan,palacios}.  The latter spectral
function is the Fourier transform of the
domain wall correlation function
$<\psi(y,\tau)\psi^{\dag}(y,0)> \sim 1/\tau^{\kappa}.$  The power law dependence
of this Green's function leads to an anomalous power law $I-V$, a
well-known signature of Luttinger liquids.  It should be
emphasized that this differs considerably from edge
state tunneling in standard 2DEGs, where the exponent is set
by the bulk filling factor.  The continuously varying exponent
we find for the edge of the graphene vacuum are a consequence
of the ferromagnetic nature of the undoped state, and its
unusual crossing of particle- and hole-like edge states.

In summary, we have studied the armchair edge of graphene in
its undoped state.  We demonstrated the ferromagnetic nature
of the system due to interactions, and showed how this leads
to a coherent domain wall due to a crossing of the non-interacting
edge states.  The domain wall can be probed by coupling it
to a normal lead, and we found this supports a power-law
tunneling current and the possibility of a quantum phase
transition into a metallic state.

{\it Acknowledgements} -- The authors thank F. Guinea and C. Tejedor for
useful discussions.  This work was supported by MAT2005-07369-C03-03
(Spain) (LB) and by the NSF through Grant No. DMR-0454699 (HAF).



\begin{thebibliography}{99}
\bibitem{com} The states are precisely at zero energy only for vanishing Zeeman coupling.
\end{thebibliography}
\end{document}